\documentclass{eas}
\usepackage{graphicx}
\usepackage{natbib}
\begin{document}

\title{Star Formation in Spiral Arms}
\author{Bruce G. Elmegreen}\address{IBM T. J. Watson Research Center, 1101 Kitchawan Road, Yorktown
Heights, New York 10598 USA, bge@us.ibm.com}
\begin{abstract}
The origin and types of spiral arms are reviewed with an emphasis on
the connections between these arms and star formation.  Flocculent
spiral arms are most likely the result of transient instabilities in
the gas that promote dense cloud formation, star formation, and
generate turbulence. Long irregular spiral arms are usually initiated
by gravitational instabilities in the stars, with the gas contributing
to and following these instabilities, and star formation in the gas.
Global spiral arms triggered by global perturbations, such as a galaxy
interaction, can be wavemodes with wave reflection in the inner
regions. They might grow and dominate the disk for several rotations
before degenerating into higher-order modes by non-linear effects.
Interstellar gas flows through these global arms, and through the more
transient stellar spiral arms as well, where it can reach a high
density and low shear, thereby promoting self-gravitational
instabilities. The result is the formation of giant spiral arm cloud
complexes, in which dense molecular clouds form and turn into stars.
The molecular envelops and debris from these clouds appear to survive
and drift through the interarm regions for a long time, possibly 100
Myr or more, with lingering spontaneous star formation and triggered
star formation in the pieces that are still at high-pressure edges near
older HII regions.
\end{abstract}
\maketitle

\section{Introduction}

An important feature of many disk galaxies is their spiral structure,
which, for the Milky Way, has been connected with star formation since
\cite{morgan53} found concentrations of OB stars in the Sagittarius
spiral arm of the Milky Way. This connection suggests that spiral arms
trigger star formation, which makes us wonder how such triggering might
fit in with the global star formation laws discussed in the previous
lecture. The answer is that spiral arms have very little influence on
large-scale star formation rates, but they do organize the star
formation in a galaxy. This is probably because most of the gas that
{\it can} turn molecular, i.e., inside shielded regions at normal
interstellar pressure, has already done so in the main disks of
galaxies, and because this molecular gas is already forming stars as
fast as it can. Star formation in this case may be viewed as saturated
\citep{e02}. Additional cloud collisions in spiral shocks, or new cloud
formation in spiral arms, does not add much to the molecular mass and
star formation-- it only moves it around. This may not be true in the
outer parts of galaxies, where the gas is highly atomic. There,
dynamical processes such as spiral arms could affect the average star
formation rate. There is very little known about outer disks yet, so
the influence of outer spiral arms on average star formation rates
remains an open question.

In this lecture, we begin with a description of spiral waves and the
various theories for them. Then we discuss detailed models for how
spiral arms interact with the gas and affect the formation of giant
clouds. We also discuss the interarm clouds and the apparent aging and
destruction of dense clouds as they move to the next arm.

\section{Spiral Waves and Modes}

Bertil Lindblad (1962) noticed that $\Omega-\kappa/2$ for angular
rotation rate $\Omega$ and epicyclic frequency $\kappa$ was about
constant with radius in galaxy disks. He suggested that spirals are
fixed patterns with an angular rotation rate $\Omega-\kappa/2$, moving
through a disk of stars with a radial variation in the stellar rotation
rate $\Omega$. Thus stars and gas move through the spiral pattern. This
was the beginning of spiral density wave theory, although it was not
quite right yet. Lindblad showed rotation curves for 3 galaxies,
$\Omega$ versus $\kappa$, and a nearly constant $\Omega-\kappa/2$.
Rotation curves were difficult measurements at that time, and
derivatives in the rotation curves, as in the evaluation of $\kappa$,
were highly inaccurate.

The main problem with Lindblad's theory was that it had no forcing.
Also, $\Omega-\kappa/2$ is not quite constant.  \cite{lin64} introduced
a more dynamically correct spiral density wave theory. They realized
that $\Omega-\kappa/2$ could be forced by the spiral's gravity to a
radial-constant value, even if it was not constant from the average
rotation curve. Then the stellar orbits could be closed for a wide
range of radii at a fixed pattern speed. The angular pattern speed
would be slightly different from $\Omega-\kappa/2$, and where it
equaled this value, there would be a resonant interaction between the
forcing from the spiral and the stellar epicyclic motions. This
resonance would absorb wave energy and put it into random stellar
motions, causing the wave to stop propagating at this place. This
position became known as the inner Lindblad resonance. Another
resonance position is where $\Omega+\kappa/2$ equals the spiral pattern
speed. This is the outer Lindblad resonance. Other resonances at
$\Omega-\kappa/3$ and $\Omega+\kappa/3$, occur as well, limiting the
range for three-arm spirals in this case. There are similar limits for
4 arm spirals, etc., and finally the last resonance where $\Omega$
itself equals the pattern speed. This is the corotation resonance,
where the same stars are always inside the wave crest, following it
around at the same angular speed.

The Lin-Shu mechanism works because just inside an arm, spiral gravity
pulls a star outward for short time, slowing it down a little as it
rises in its epicyclic path. Just outside an arm, spiral gravity pulls
the star inward, speeding it up as it falls inward in its epicyclic
motion.  These slow-downs and speed-ups cause the ends points of each
epicycle to advance a little, closing the orbits in a rotating frame
with a rate $\Omega_p$ such that $\Omega-\kappa/2 < \Omega_p <
\Omega+\kappa/2$. The gravitational effect can be seen in the
dispersion relation written by \cite{toomre69}:
\begin{equation}
\left(\omega-m\Omega[r]\right)^2=\kappa^2(r)-2\pi G\Sigma(r)k{\cal
F}(\chi)\end{equation} where $\omega$ is the rate of change of the
spiral phase in a fixed coordinate system, equal to $m$ times the
pattern speed, $m$ is the number of symmetric spiral arms, $\Sigma$ is
the mass column density in the disk, $k$ is the wavenumber, and ${\cal
F}$ is an integral over stellar motions that depends on
$\chi=k^2\sigma_{\rm u}^2/\kappa^2$ for rms radial speed of the stars
$\sigma_{\rm u}$. Lindblad's theory did not have the last term on the
right, which is from disk gravity. Traveling waves exist for Toomre
parameter $Q>1$, i.e., for disks that are stable to radial
perturbations.

\cite{toomre69} noted that although the Lin-Shu dispersion relation for
spiral waves has a phase velocity equal to the proposed pattern speed,
it also has a group velocity which causes the wave crests to move
inward, i.e., the spirals wrap up. Thus the ``quasi-stationary'' spiral
density wave theory of Lin, Shu, Roberts, Yuan, and other collaborators
at that time, did not work as they originally proposed. Toomre showed
that for a flat rotation curve, disk-dominated gravity, and constant
stability parameter $Q$, the Lin-Shu dispersion relation becomes
relatively simple,
\begin{equation}
\left(\omega-m\Omega\right)/\kappa = \left(m/2^{1/2}\right)\left[
\left(r/r_{\rm CR}\right)-1\right]\end{equation} for radius $r$ and
corotation radius $r_{\rm CR}$. In this case, the time derivative of
the dimensionless wavenumber increases at a rate equal to half the rate
of change in the phase, $\omega/2$. When the wavenumber increases with
time, the spirals get closer together, which means they migrate inward.
This is a fast migration, almost as fast as purely material arms would
wrap up from shear.

\cite{toomre69} proposed that spirals are not quasi-stationary, but
transient, provoked either by interactions \citep{tt72} or noise
\citep{tk91}. \cite{kn79} noted that ``grand design'' spirals are
either in barred galaxies, in the rising parts of rotation curves
(where $\Omega-\kappa/2\sim0$) or in interacting galaxies. This would
be consistent with Toomre's picture. \cite{toomre81} identified the
cause of transient spirals as ``swing amplified instabilities.'' Many
groups have studied these instabilities numerically
\citep[e.g.,][]{fuchs05}.

The quest for a theory of quasi-stationary spiral structure was not
over, though. \cite{mark74}, \cite{lau76}, and \cite{bertin89} proposed
a ``modal theory'' in which inward-moving waves reflect or refract off
of a bulge or bar and come back out as leading (WASER2; reflection) or
trailing (WASER1; refraction) spiral arms. When they reach the
corotation resonance moving outward, they amplify. Part of the wave
then turns around to come back in and another part of the wave keeps
going outward. The result is a standing wave pattern, amplified from
initial disk noise at corotation and forming a long-lived grand-design
spiral. The corotation radius is where the outward-moving wave meets
the inward-moving wave on the opposite side of the galaxy for a two-arm
spiral. If the outward moving wave is leading, then at the meeting
place, the swing amplifier can transform this leading wave into a
strong trailing wave. For example, an inward moving wave starting at
corotation in one arm of a two-arm spiral can reflect off of a bulge
and move back out as a leading wave until it meets the other arm at the
same radius where it started. It amplifies as it is converted into a
trailing arm, adds to the original trailing arm, and then a stronger
trailing arm comes in again. Trailing waves that start at different
radii will not reflect and meet the opposite arm at the same radius,
and so will not add to the original wave after amplification. Thus, out
of all the disk noise and small spirals that they initiate, only the
spiral with the ability to amplify reflected or refracted waves and
reinforce itself will grow. This defines the corotation radius.
\cite{bertin89} described this process in detail.

Spiral wave modes could exhibit an interference pattern between the
inward and outward moving waves. Interference acts to modulate the
amplitudes of the main arms or it may introduce slight phase shifts in
the main arms. Such modulation is present in the model solutions shown
by \cite{bertin89}. \cite{eem92} reported such interference patterns,
but a more modern analysis is needed.

The various theories of spiral wave formation may be reduced to four
basic types: random and localized swing-amplified spirals that are
primarily in the gas (because the stellar disk is somewhat stable);
random and localized swing-amplified spirals that are in the stars and
the gas together; transient global waves that are in the stars and gas,
and standing wavemodes that are in the stars and gas. The first type
produces flocculent spiral arms and a smooth underlying stellar disk
(e.g., NGC 5055), the second type produces multiple stellar and gaseous
arms (e.g., NGC 3184), the third type produces long spiral arms in the
stars and gas (e.g., NGC 628), and the fourth type produces strong
two-arm spirals in the stars and gas, usually in response to some
global perturbation like a galaxy interaction (e.g., M51, M81). Aside
from M81, these galaxy examples were chosen from the THINGS survey
\citep{walter08}.

A spiral wavemode may be compared to the pure-tone ringing of a bell
after some multi-frequency impact disturbs it (e.g., it is hit by a
hammer or bowed by a violin string).  Random swing-amplified spirals
have been called spiral chaos. They are the primary response to
gravitational instabilities in the stars and gas and therefore have a
strong connection with star formation and the origin of interstellar
turbulence in the absence of global wavemodes
\citep[e.g.,][]{thomasson92, bournaud10}.  Global spiral waves or
wavemodes also have a connection with star formation because of the way
they force the gas into a dense molecular phase in the dust lanes
(which are shock fronts) and organize it to follow the underlying
stellar spiral.

\section{Motions in Spiral Arms}

Because of the forcing from gravity, spiral arms induce a reverse shear
in the stellar rotation, slowing down the stars on the inner parts and
speeding up the stars on the outer parts of each arm \cite{roberts69}.
This reverse often cancels the normal shear from average orbital
motions, and makes an arm that has very little internal shear. The arm
forcing from gravity also pulls everything in the arm toward the center
of the arm, i.e., in a convergent manner, which is opposite to the
tidal force from the surrounding galaxy. Thus spiral arms also have
reduced galactic tidal forces \citep{e92}. These conditions are good
for the formation of large cloud complexes, which are only weakly bound
at the start. Giant clouds that form by gravitational instabilities in
spiral arms do not immediately shear out into little spirals, and this
allows them to grow. When they emerge from the arm, the shear rate and
tidal disruption rate increase a lot, and the low-density parts of the
clouds can come apart. They can also form feathers and spurs. Such
feathering is commonly seen \citep{lavigne06}. Feathers occur primarily
in grand design (2-arm) galaxies with prominent dust lanes. The
feathers are closer where gas density is highest, as expected for
gravitational instabilities. Their separation is $5-10$ Jeans lengths
in the dust lane \citep{lavigne06}.

The peculiar motions from spiral arm gravity and Coriolis forces cause
the gas and stars to stream along the spiral arms when they are in the
arms, and to expand away from the arms when they are between the arms.
Streaming motion of the gas can be very strong, perhaps 50 km s$^{-1}$
or more, as in M51 \citep{shetty07}. Radial streaming changes sign at
corotation and the observation of this allows one to locate the
corotation resonance radius \citep[e.g.,][]{ewp98}. Streaming motions
also allow one to measure the timing of the star formation response to
the spiral arm \citep{tamburro08}. The streaming pattern for gas tends
to be inward inside the main parts of the arms inside the corotation
radius, and outward in the interarms inside corotation. This pattern of
radial motions relative to the arms reverses outside of corotation.
Thus the gas and star formation in outer disk spirals, as viewed, for
example, by GALEX, should be streaming with positive galacto-centric
radial velocities, whereas the gas and star formation in inner disk
spirals should be streaming with negative galacto-centric radial
velocities. \cite{shetty07} found a net radial inward streaming flux
for the inner part of M51, including both the arms and the interarms.
They suggested that this meant a non-steady spiral pattern.

\section{Magnetic Fields in Spiral Arms}

The magnetic fields of spiral galaxies have been extensively observed,
particular by Rainer Beck and collaborators \citep[e.g.,][]{braun10}
using Faraday rotation. In NGC 6946 \citep{beck07}, the field structure
is uniform in the interarm regions, probably from the combing action of
shear with little disruption from star formation. It is more chaotic in
the arms, and even weaker on average in the arms than the interarms
because of the strong random component in the arms. The total field
strength in the arms should be higher than in the interarms because of
compression from the spirals, but this higher field strength might not
be seen with rotation measures if the field direction fluctuates on
small scales.

Gravitational instabilities in spiral arm gas are enhanced by the lack
of shear and the magnetic field \citep{e87}, which tends to run
parallel to the arms in the direction of the unstable flow. This field
removes angular momentum from a growing condensation, as mentioned in
Lecture 1. \cite{ko01} have dubbed this the Magneto-Jeans instability,
and modeled it numerically. Gas collapses along spiral arms into giant
cloud complexes, aided by the parallel magnetic field. The dense gas
then emerges downstream from the arms as interarm feathers.

\section{Gravitational Instabilities in Spiral Arms}

Gas rich galaxies with weak stellar spirals have more interarm gas and
star formation than galaxies with strong stellar spirals. In weak-arm
spiral galaxies, local swing-amplified instabilities in the gas become
prominent and these can occur almost anywhere. In galaxies with strong,
global stellar waves, the magneto-Jeans instability forms giant cloud
complexes primarily in the spiral arms, where the density is high and
the shear is low.

\begin{figure}[b]
\begin{center}
 \includegraphics[width=3.in]{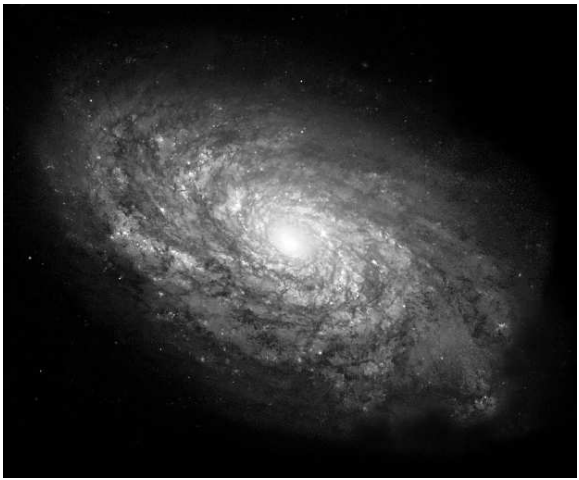}
\caption{Hubble Space Telescope image of NGC 4414, from multiple
passbands.}\label{f1}
\end{center}
\end{figure}

\begin{figure}[b]
\begin{center}
 \includegraphics[width=3.in]{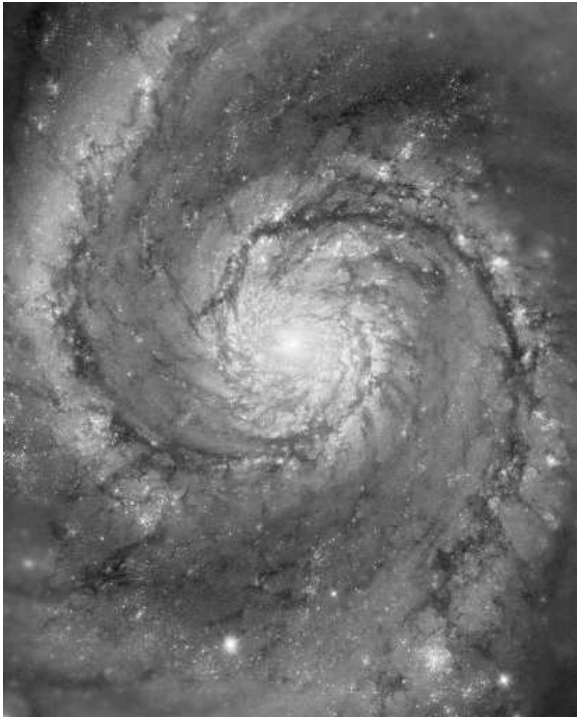}
\caption{Hubble Space Telescope image of NGC 5194, from multiple
passbands.}\label{f2}
\end{center}
\end{figure}

Figure \ref{f1} shows a Hubble Space Telescope image of the galaxy NGC
4414, which has numerous patches of star formation in the midst of a
faint 2-arm structure.  This is an example of the first type mentioned
in the previous paragraph. Figure \ref{f2} shows an HST image of M51, a
strong two-arm spiral with little star formation between the stellar
arms. This is an example of the second type.

\begin{figure}[b]
\begin{center}
 \includegraphics[width=3.in]{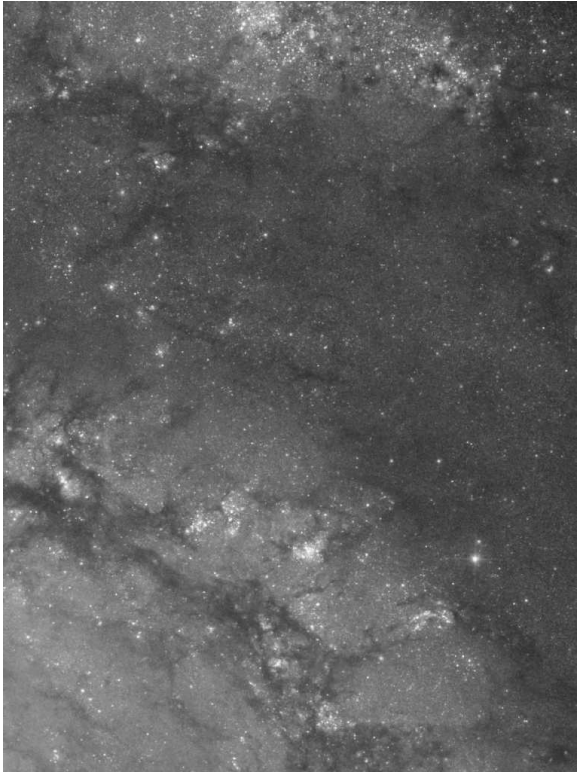}
\caption{Enlargement of the Western interarm region of M51, from the
Hubble Space Telescope image. Dark dust clouds with small amounts of
star formation, or no evident star formation, are seen. Some are at the
edges of old OB associations and may contain triggered star formation.
}\label{f3}
\end{center}
\end{figure}

Stellar spirals define two scales, $2\pi G\Sigma/\kappa^2$, which is
the ``Toomre length'' \citep{toomre64} separating the spirals, and
$2\sigma^2/G\Sigma$, which is the Jeans length separating the
condensations in the spirals. The Jeans length is about three times the
arm width and physically smaller inside the dense dust lanes. The beads
on a string seen in spiral arms are giant star complexes. Each has a
feather or spur of dust from a spiral wave flow downstream. This is
clearly visible in the HST image of M51, shown in Figure \ref{f2}. In
the interarm regions of M51, young stars are still in star complexes
that are aging. Lingering star formation and triggered star formation
occur in the interarm fields of cloudy debris. Further downstream, the
cloud envelopes are more diffuse but there is still a little star
formation in some of them (Fig. \ref{f3}). The molecular envelopes of
GMCs must be long-lived to survive as far as they do downstream from
the arms, $\sim100$ Myr or more. This seems to require magnetic support
in the cloud envelopes \citep{e07}. Further downstream, almost at the
next spiral arm, the cloudy debris from the previous arm has little
associated star formation. There appears to be a lot of diffuse
molecular gas indicated by these interarm dust features. They coagulate
into a dust lane when they reach the next arm.

At a very basic level, a gravitational instability in a spiral arm, or
in a spiral arm dust lane \citep{e79} can be viewed as an instability
in a cylinder. Such instabilities occur when
\begin{equation}
\pi G \mu /\sigma^2 \geq 1\end{equation} where $\mu$ is the mass per
unit length in the cylinder and $\sigma$ is the velocity dispersion.
The fastest growing mode has a wavelength of about 3 times the cylinder
width \citep{ee83}.

There is also enhanced star formation at the end of some strong bars.
This is most likely a crowding effect from the gas that turns a corner
there in its orbit relative to the bar \citep{lk91}. The inner, nearly
straight, dust lanes in many bars do not contain much star formation
and look non-self gravitating. This is probably because of high shear
and radial tidal forces. Inside this dustlane, in the center of the
bar, there is often a ring close to the inner Lindblad resonance
\citep{bc96}. This ring also has two characteristic scales, the
thickness in the radial direction and the Jeans length. ILR rings
develop major sites of star formation, or ``hot-spots'' along them,
with a separation of around the Jeans length, probably because of local
gravitational instabilities in the gas \citep[e.g.,][]{e94}.

\section{Star Formation in Spiral Arms}

What is the relationship between HI, CO and star formation in spiral
arms? The gas is generally compressed more than the stars in a spiral
density wave or swing-amplified transient spiral, and if star formation
follows the gas, then the blue light from star formation will be
enhanced more than the yellow and red light from old stars. This makes
the spirals arms blue. The \cite{bigiel08} and \cite{leroy08}
correlation between SF rate and CO, which is a very tight correlation,
implies that there is little difference in the rate per unit CO
molecule for gas in strong-arm galaxies compared to gas in weak-arm
galaxies.

The morphology of gas in the arms tells something about the star
formation process.  \cite{grabelsky87} showed that most of the CO
clouds in the Carina arm of the Milky Way are clustered together in the
cores of $10^7\;M_\odot$ HI clouds. \cite{ee87} found the same for the
Sagittarius spiral arm. \cite{lada88} observed a similar HI envelope-CO
core structure in a piece of a spiral arm in M31. \cite{engargiola03}
showed a complete map of M33 with numerous CO clouds in the cores of
giant HI clouds. The presence of giant HI clouds in spiral arms has
been known for a long time (e.g., for the Milky Way: McGee \& Milton
1964; for NGC 6946: Boulanger \& Viallefond 1992). Now it looks like
most dense molecular clouds are in the cores of giant spiral arm HI
clouds, or if the gas is highly molecular at that radius in the disk,
in the cores of giant clouds that are also highly molecular. This means
that GMCs form by condensation inside even larger, lower-density
clouds. The large HI/CO clouds, in turn, probably form by gravitational
instabilities in the spiral arm gas, particularly in the dust lanes
where the spiral shock brings the gas to a high density. Recall from
Lecture 1 that the largest unstable clouds have the Jeans mass in a
galaxy disk, given the observed turbulent speed and column density
(i.e., $M\sim\sigma^4/G^2\Sigma_{\rm gas}\sim 10^7\;M_\odot$).

In the Milky Way and M33, giant spiral arm clouds are mostly atomic,
but in M51, they are mostly molecular \citep{rk90}. This difference is
presumably because the arms in M51 are much stronger than the arms in
the Milky Way and M33, and the gas is denser overall in M51 as well.
Thus, the pressure is higher in M51, particularly in the arms, and the
gas is more highly molecular there and everywhere else in the inner
disk. The physical process of giant cloud formation should be the same
in all three cases, however.

Gravitational instabilities also seem to initiate cloud and star
formation on the scale of whole galaxies. This process is clear in many
regions, such as Stephan's quintet \citep{mendes04}, NGC 4650
\citep{karataeva04}, and in the tidal arcs of NGC 5291 (Bournaud et al.
2007), where there are massive condensations in tidal features.

\cite{dp09} studied gravitationally bound clouds in an SPH simulation.
In the spiral arms, large regions formed by gravitational instabilities
where gravity balanced thermal, turbulent and magnetic energies. When
they used a star formation rate equal to 5\% of the bound gas column
density divided by the dynamical time, plotted versus the total gas
column density, they reproduced the \cite{ken98} and \cite{bigiel08}
star formation laws over the range of overlap. They noted that the star
formation law is linear with column density because the dynamical time
inside each bound cloud is the same, i.e., they all have the same
density. They answered the long-time question of whether density waves
trigger star formation \citep{ee86} by saying, no, there is no
correlation between the average column density of star formation and
the spiral arm potential depth. The reason is that stronger spiral
waves make clouds with higher velocity dispersions and they are harder
to bind into gravitating cloud complexes.  The fraction of the bound
gas in spiral arms increases with the spiral strength, but not the star
formation rate.

Observations of spiral arm star formation and gas distributions also
suggest there is little triggering \citep{foyle10}. The primary effect
of the spiral is to concentrate the gas in the arms without
significantly changing the star formation rate per unit gas.

\section{ Summary}

Spirals can, in principle, be of 4 types: (1) Transient gravitational
instabilities in the gas, causing ``flocculent spirals,'' with too much
stability in the stellar disk to give prominent stellar spiral waves.
(2) Transient gravitational instabilities in the stars, with the gas
adding force and following the stars. The gas and stars move through
these transient spirals a little, but not around from arm to arm in a
full circle as in idealized global stellar modes and waves. (3) Global
stellar waves that are non-steady with a pattern speed that varies with
radius and whose patterns wrap up toward the center over time. Stars
and gas move through these spirals. (4) Global stellar wave modes that
are ``standing waves,'' with a uniform pattern speed between the
Lindblad resonances.  Gas and stars move through these standing waves
with corotation approximately at mid-radius in the spiral.

Young stars concentrate in spiral arms because the gas concentrates
there. Spiral arms are dense and promote more gravitational
instabilities and cloud collisions than the interarm regions,
triggering molecular cloud formation and conglomeration in the arms.
The star formation rate per unit area is high in the arms as a result.
This excess star formation rate is mostly in proportion to the extra
molecular gas column density there, without a significant change in the
star formation rate per unit molecular gas mass. The total galactic
star formation rate in the main disk is not significantly enhanced by
the presence of spiral arms. That is like saying the gas would have
formed the same abundance of molecular clouds even without the arms.
Outer disks may be different. They may have an excess of total star
formation if there are spiral arms there, but this excess has not been
observed yet.  The difference between inner disks and outer disks is
that inner disks are highly molecular and star formation in the gas is
virtually saturated. Outer disks are mostly atomic and without star
formation, so triggering a higher rate of star formation might be
possible with dynamical disturbances.


\end{document}